\documentclass[sigconf]{acmart}

\PassOptionsToPackage{sort&compress}{natbib}
\usepackage{natbib}

\copyrightyear{2025}
\acmYear{2025}
\setcopyright{cc}
\setcctype{by}
\acmConference[CF '25]{22nd ACM International Conference on Computing
Frontiers}{May 28--30, 2025}{Cagliari, Italy}
\acmBooktitle{22nd ACM International Conference on Computing Frontiers (CF
'25), May 28--30, 2025, Cagliari, Italy}\acmDOI{10.1145/3719276.3725174}
\acmISBN{979-8-4007-1528-0/2025/05}

\begin{document}

\title{FERIVer: An FPGA-assisted Emulated Framework for RTL Verification of RISC-V Processors}

\author{Kun Qin}
\email{kun.qin@tum.de}
\affiliation{%
  \institution{Technical University of Munich}
  \city{Heilbronn}
  \country{Germany}
}

\author{Xiaorang Guo}
\email{xiaorang.guo@tum.de}
\affiliation{%
  \institution{Technical University of Munich}
  \city{Garching}
  \country{Germany}
}

\author{Martin Schulz}
\email{schulzm@in.tum.de}
\affiliation{%
  \institution{Technical University of Munich}
  \city{Garching}
  \country{Germany}
}

\author{Carsten Trinitis}
\email{carsten.trinitis@tum.de}
\affiliation{%
  \institution{Technical University of Munich}
  \city{Garching}
  \country{Germany}
}

\renewcommand{\shortauthors}{Qin et al.}


\begin{abstract}
Processor design and verification require a synergistic approach that combines instruction-level functional simulations with precise hardware emulations. The trade-off between speed and accuracy in the instruction set simulation poses a significant challenge to the efficiency of processor verification. By tapping the potentials of Field Programmable Gate Arrays (FPGAs), we propose an FPGA-assisted System-on-Chip (SoC) platform that facilitates cross-verification by the embedded CPU and the synthesized hardware in the programmable fabrics. This method accelerates the verification of the RISC-V Instruction Set Architecture (ISA) processor at a speed of 5 million instructions per second (MIPS), which is 150x faster than the vendor-specific tool (Xilinx XSim) and a 35x boost to the state-of-the-art open-source verification setup (Verilator). With less than 7\% hardware occupation on Zynq 7000 FPGA, the proposed framework enables flexible verification with high time and cost efficiency for exploring RISC-V instruction set architectures.
\end{abstract}







\maketitle

\section{Introduction}
The evolution of modern processor architectures and Instruction Set Architectures (ISA) demands efficient architectural modeling and hardware verification techniques. Although the principles discussed may have broader applications on ARM and X86 architectures driven by the integration of edge computing, energy-efficient designs, and quantum computing innovations \cite{hisepq}, this paper focuses on the solutions within the context of RISC-V due to its flexibility and openness, which allow ISA extensions and integration with customized accelerators. The growing complexity of RISC-V processors underscores the need for advanced verification platforms to ensure reliability and functionality in diverse implementations. Conventional simulation and verification take most of the time and require huge financial and power inputs in processor verification. Specifically, more than 60\% of the total efforts are consumed by verification during development iterations~\cite{akram_survey_2019}. Therefore, potential benefits in terms of accuracy and efficiency are gaining increased interest in verification methodologies. 

Instruction-set Simulators (ISSs) are tools for abstractly simulating specific processor instruction sets without physical hardware. By providing a virtual environment, ISS facilitates software debugging, performance analysis, architectural exploration, and design validation across various processor architectures. When developing new RISC-V products, open-source ISSs are indispensable tools since they serve as reference models for rigorous hardware verification, ensuring functional correctness. Furthermore, open-source ISSs are invaluable for exploring architectural concepts and performance optimization techniques, contributing to the cost-effective development and widespread adoption of increasingly used RISC-V, which makes it a compelling choice for diverse applications.


From the development platform perspective, the advancements in modern Field Programmable Gate Array (FPGA) technologies enable FPGA-based platforms to significantly enhance the design verification process~\cite{statemover_2020}.
An FPGA System-on-Chip (SoC), combining a Processing System (PS) with Programmable Logic (PL), offers significant advantages over traditional FPGAs by integrating software programmability and hardware acceleration on a single chip. 
This tight integration reduces latency, improves power efficiency, and allows high-speed communication between PS and PL via internal buses. FPGA SoCs also simplify system design with software/hardware co-development tools and reduce the physical footprint by eliminating external components. FPGA SoCs provide a versatile, cost-effective solution for heterogeneous architectures, combining the flexibility of software with the performance of hardware.

In this case, by leveraging both ISS and FPGA SoC technologies in tandem, execution discrepancies of a processor prototype can be identified through a systematic comparison between 1) the abstract architectural model provided by the ISS and 2) the actual implementation in the programmable logic area of FPGAs. However, adopting this agile method to verify the new computer architecture comes with the following challenges:
\begin{itemize}
    \item \textbf{Selection of ISS: fine-grained simulation results in low speed}. For example, the timing-accurate ISS or the Register-transition Level (RTL) simulation matches the exact timing of each hardware component but is extremely slow \cite{gvsoc}. In contrast, the function-accurate ISS focuses on the behavior of processor instructions without modeling internal hardware details, which is preferable for speed with little detail about the micro-architecture of the processors. Therefore, the ISA simulation and verification inherently present a trade-off between granularity and speed. To address this trade-off, the coarse-grained verification of function-accurate ISS must be compensated with more detailed and precise hardware emulation, such as RTL emulation on FPGA, to ensure comprehensive validation.
    \item \textbf{Debugging Interface on FPGA: signal-monitoring of complex RTL designs demands significant hardware resources}. The emulation of RTL designs on FPGA is a promising solution to compensate for the function-accurate ISS, while it requires debugging ports for tracking the signals, through which the host machine monitors the transaction of internal signals within the programmable area. Customized logic is able to extract signals from the designed entities, while finite FPGA resources limit the number of signal probes. Moreover, the communication overhead between the host and programmable fabric ultimately constrains high-speed verification at runtime. Thus, establishing an efficient and scalable debugging interface between the host machine and the FPGA remains a significant challenge. 
\end{itemize}

To address these challenges, we propose the FERIVer framework, which employs the function-accurate ISS and composes suitable driver interfaces to parse the checkpoints of the hardware emulation on FPGA. A dumped waveform of the checkpoint is generated when the ISS and PL have different execution results. As such, the co-verification method meets the need for a higher speed, and the accuracy loss of ISS can be compensated by RTL emulation on FPGA.

The remainder of this paper is structured as follows. In \hyperref[sec-back]{\S 2}, we present the background by examining the context of ISA simulation and FPGA-based processor verification technologies. \hyperref[sec-frame]{\S 3} introduces our FERIVer framework, providing an overview of its operational principles and critical design decisions. In \hyperref[sec-evaluation]{\S 4}, we present a comprehensive evaluation of our verification methodology and compare its performance against existing solutions. The paper concludes in \hyperref[sec-conclusion]{\S 5} with a summary of our contributions and potential avenues for future research in this domain.

\section{Background} \label{sec-back}
Processor verification requires a combination of instruction-level simulation and hardware-level emulation. To ensure the fidelity of the actual micro-architecture, this dual-pronged approach is crucial to compare the ISS execution results with those obtained from hardware emulation, verifying the functional correctness and helping in timing analysis. This comparison process allows designers to validate that the hardware implementation accurately reflects the architecture of the intended instruction set.
Commercial application-specific design tools, such as Synopsys\textsuperscript{\textregistered} ASIP \cite{SynopASIP} and Cadence\textsuperscript{\textregistered} Tensilica \cite{CadenTie}, provide features like ISA exploration and micro-architecture synthesis. For high-end processor verification, some off-the-shelf hardware emulators such as Siemens\textsuperscript{\textregistered} Veloce Strato \cite{MentorVel}, Synopsys\textsuperscript{\textregistered} ZeBu EP2 \cite{SynopZeBu}, enable emulation with real-world environments to uncover bugs and hidden issues. Not to mention their expensive license fees and significant machine costs \cite{MentorVel} \cite{SynopZeBu}, they do not provide alternatives for small groups of researchers or developers in the open-source community.

\subsection{Open-source Software-based Processor Simulator}
Software-based ISS provides an efficient ISA simulation, offering a high-level model of the processor's behavior. ISS can be classified based on the axes of accuracy or specialized architectures \cite{akram_survey_2019}. This research focuses exclusively on open-source ISS that support RISC-V ISA, and we study ISS categories on the first axis of accuracy: timing-accurate ISS and function-accurate ISS.

Timing-accurate simulators offer a fine-grained abstraction of the micro-architecture, encompassing cycle-accurate, instruction-execution-driven, and event-driven categories. Cycle-accurate simulators deal with the RTL design, such as Verilator \cite{verilator}, providing the highest accuracy but taking a longer time, typically achieving a performance of 0.01$\sim$0.1 million instructions per second (MIPS) \cite{verilator_doc}. Instruction-execution-driven simulators, exemplified by Gem5 (MinorCPU mode) \cite{gem5}, strike a balance between timing accuracy and verification speed (0.1$\sim$1 MIPS) using partial mapping of the target micro-architecture. At a higher level of abstraction, event-driven simulators (e.g., RISC-V TLM \cite{Monton2020ARS} or GvSoc \cite{gvsoc}) offer superior speed improvements (1$\sim$100 MIPS) based on Transaction Level Modeling (TLM) \cite{gvsoc}. In essence, the simulation speed of timing-accurate ISS is impeded by the computational overhead of detailed modeling of the low-level micro-architecture. In contrast, function-accurate ISS prioritizes speed over timing accuracy. These simulators, such as Spike \cite{spike} and QEMU \cite{qemu}, replicate functional behavior without modeling the architectural sub-modules, thereby sacrificing micro-architectural insights in favor of rapid simulation in kMIPS\cite{Monton2020ARS}. To compensate for the drawback of function-accurate ISS, FPGA emulation is introduced to provide extra timing accuracy during the verification of processor architectures \cite{wang_fpga_2010}.

\subsection{Verification on Heterogeneous FPGA SoC}
The effectiveness and re-usability across multiple projects make FPGA platforms highly cost-efficient. To verify a processor prototype, developers can choose between RTL simulation solely on the CPU and emulating the processor in the programmable area at run time. The former method possesses full-scale visibility at an extremely slow rate, and the latter runs with fewer signal probes but at a higher speed, around 100,000x compared to CPU-based RTL simulation \cite{statemover_2020}.

The method of monitoring the signals in the FPGA-based systems varies significantly across different platforms. For example, XtremeData\textsuperscript{\textregistered} FPGA, which seamlessly integrates into Intel\textsuperscript{\textregistered} Xeon\textsuperscript{\textregistered} server sockets, offers comprehensive signal extraction capabilities from the programmable area to the hosting CPU \cite{wang_fpga_2010}. However, signal monitoring in universal consumer-grade FPGAs presents more complex and cost-intensive challenges. On the vendor-defined FPGA SoC, developers need customized components to observe specific signals of interest.


\subsubsection{\textbf{\textit{Manually Created RTL Signal Probes}}}
In \cite{trace}, authors propose a trace-based method that stores the marked signals in on-chip block RAM (BRAM) upon the trigger conditions, where the host can access and process the data. Hence, the continuous data streams are fed into the dedicated signal buffers, thus consuming more precious BRAM and limiting the number of monitored signals. An optimal scan-based solution was published in DESSERT \cite{scan}. It compiles a scan chain in the device under test (DUT) and compares it with another reference instance that is also implemented in the hardware, i.e., one signal is captured upon a mismatch between the DUT and the golden reference. Assuming a finite memory volume, the scan-based method can record more useful data intermittently. Nonetheless, the overhead of the larger "shadow" instance becomes dominant with the increasing size of the DUT. In conclusion, both methods are customized circuits and sacrifice hardware utilization for speedy verification in PL.

\subsubsection{\textbf{\textit{Signal Analyzer IP}}}
To meet the needs of high-speed simulation and more signal probes, vendors provide predefined intellectual property (IP) modules for signal surveillance and verification at PL runtime, such as AMD\textsuperscript{\textregistered} Integrated Logic Analyzer (ILA) \cite{amdILA} and ChipScope~\cite{amdchipscope}. Those vendor-specific signal tracking techniques request that users specify the signals of interest together with the design entity. This rigid approach necessitates recompilation of the design whenever additional signals are required for monitoring. Furthermore, ILA's reliance on on-chip memory for data buffering can impose significant area overhead on the FPGA fabric. They are more efficient than manually implemented probe circuits due to the optimized logic and on-chip memory usage. Nevertheless, they still require extra time for recompilation when the probing logic is modified, and more hardware utilization, placement, and routing complexity are added to the implementation, limiting the temporal sampling depth of the signals.

\subsubsection{\textbf{\textit{Bitstream Readback}}} \label{R6}
In addition to inserting signal probes in the RTL design, partial reconfiguration is another FPGA technology provided by AMD\textsuperscript{\textregistered} Xilinx Zynq/UltraScale \cite{amdILA} and Intel\textsuperscript{\textregistered} Stratix 10 FPGAs \cite{intelstratix}, namely, bitstream "Readback". The readback of certain parts of the bitstream in advanced FPGA SoC facilitates innovative interfaces designed to enhance PL configuration capabilities. For example, the Processor Configuration Access Port (PCAP) \cite{PCAP} in Xilinx FPGA SoCs provides a unique mechanism for reading data frames from the deployed bitstream. 


On AMD/Xilinx Zynq-7000 series FPGA SoC, each frame within the deployed bitstream is assigned a unique Frame Address (FRAD) that specifies its location on PL. The processing system (ARM) accesses frames sequentially, starting with the frame indicated by the current value of the Frame Address Register (FAR) that contains the FARD. The FAR automatically increments to the next frame address after each read operation unless the end of a row is reached. The number of frames read during a readback operation is precisely determined by the number of words specified in the read request, taking into account any necessary padding frames for recognition.

The Frame Data Register Out (FDRO) serves as the output port for the configuration memory. Each data word read from the FDRO is directly sourced from the on-chip memory at the frame address currently held by the FAR. Prior to any read operation, the number of words to be transferred must be explicitly written to the appropriate register.

The PCAP readback mechanism used in this work is discussed in \hyperref[sec-frame]{\S 3.1.2}. 


\begin{figure*}[bpht]
\centerline{\includegraphics[scale=0.7]{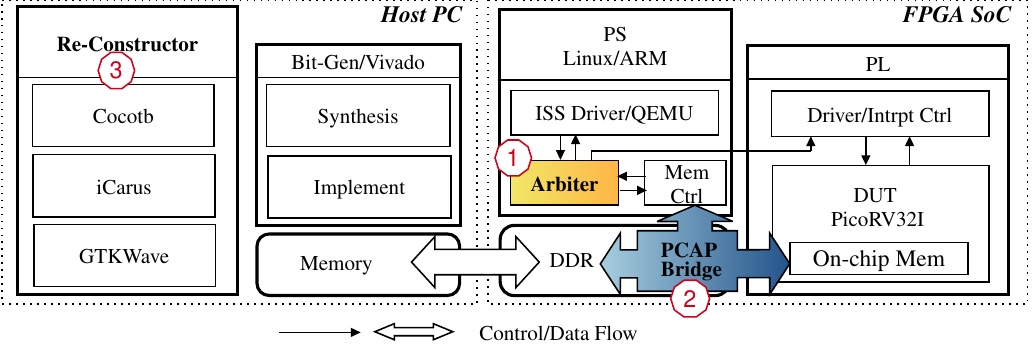}}
\caption{Block diagram of the system-level architecture.}
\Description{Communication structure between host and fpga}
\label{blockdiagram}
\end{figure*}

\section{FERIVer Framework}  \label{sec-frame}

We implement FERIVer on the AMD Zynq XC7Z020 SoC equipped with 85K programmable logic cells (Artix-7) and dual-core ARM Cortex A9 application processing units (PS), which is connected to the host PC with an AMD Ryzen 16-core processor running at 2.7 GHz. The core functionality of FERIVer lies in its capacity for cross-verification during the operation of the processor design under test. Using an FPGA SoC, FERIVer detects errors during execution and provides insights into the RTL design. This helps developers identify and fix issues more effectively.

This approach not only achieves a verification in a cycle-accurate manner by function-accurate ISS but also significantly accelerates the procedures, thus addressing the inherent speed-accuracy trade-off in modern processor design verification. 

In this section, we provide a detailed explanation of our framework. We begin by introducing the top-level architecture of FERIVer, describing its key components and their interactions. Building on this foundation, we then outline the typical workflow of the framework, demonstrating its practical application by an illustrative synthetic example. To complete our analysis, we expound upon the acceleration mechanism, illustrating how FERIVer expedites the RTL design verification process. Through this systematic exploration, we aim to showcase the advantage and efficiency of FERIVer in assisting the field of processor design verification.

\subsection{Top-level Architecture}  \label{top}



As shown in Figure \ref{blockdiagram}, the FERIVer framework enables the parallel processing of the ISS and the RTL processor model on the same FPGA SoC. The software-based ISS implementation is powered by the QEMU emulator that is running on the ARM-based processing system, while the open-source RTL RISC-V core employed is the PicoRV32I \cite{pico}. These two distinct processor representations are synchronized at the instruction level during runtime, facilitating a comprehensive and efficient co-simulation environment. This parallel execution approach allows for the seamless integration of high-level functional validation, enabled by the ISS, with the detailed architectural exploration afforded by the RTL design entity. In addition to the Vivado toolchain on the host PC, the key components of our framework are composed of:

\subsubsection{\textbf{Arbiter} (block \textcolor{red}{\textcircled{1}} in Figure \ref{blockdiagram})}
\begin{figure}[hbpt]
\centerline{\includegraphics[scale=0.7]{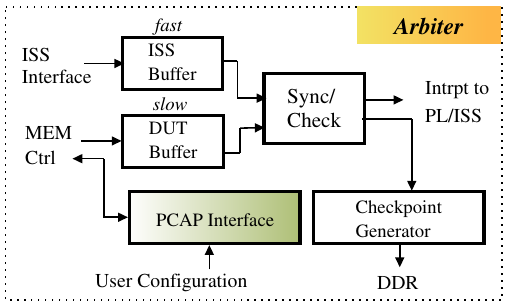}}
\caption{An Arbiter is designed for PCAP initialization and mismatch checking .}
\Description{structure of the arbiter}
\label{arbiter}
\end{figure}

The Arbiter, as shown in Figure \ref{arbiter}, checks the execution results (values in the general purpose registers) from both ISS and the RTL core implemented in PL. It generates a checkpoint upon a mismatch. Since there are two clock domains (ISS at 845MHz and PL at 100MHz), we use two buffers for receiving asynchronous data from ISS and DUT, respectively. The 'Sync/Check' unit synchronizes and compares states from two sources, parses signals for checkpoint generation, and issues interrupts to the DUT. 

\begin{table}[hbpt]
\caption{Segmentation of the frame address (Xilinx Configuration Guide UG470 \cite{amdpcap}).}
\begin{center}
\begin{tabular}{p{0.15\textwidth} | p{0.06\textwidth} | p{0.2\textwidth}}
\hline 
\textbf{Type} & \textbf{Index}& \textbf{Description}\\
\hline
Block Type & [25:23] & Valid block types. A normal bitstream doesn't include type 011. \\
\hline
Top/Bottom Bit & [22] &Selects between top-half rows (0) and bottom-half rows (1).  \\
\hline
Row Address & [21:17] & Selects the current row.  \\
\hline
Column Address 1 & [16:7] & Selects a major column, such as a column of CLBs.\\
\hline
Column Address 2 & [6:0] & Selects a frame within a major column. \\
\hline
\end{tabular}
\label{far}
\end{center}
\end{table}

The "PCAP Interface" processes the user's configuration and initializes the first frame address (FARD in FAR), which navigates the bitstream-readback transaction to the dedicated frame. As depicted in Table \ref{far}, the FRAD is composed of several segments, one of which is the block type -- determined by the highest three bits of the FRAD, which defines the specific function of the frame, such as CLK (000), Block RAM (001) and CLB (010). If monitoring new registers is necessary, designers only need to change the configuration instead of modifying the Verilog code again, saving considerable time for re-compilation and implementation.

\subsubsection{\textbf{PCAP Bridge} (block \textcolor{red}{\textcircled{2}} in Figure \ref{blockdiagram})}
The customized PCAP data path typically begins with the processing system issuing a readback request through the "PCAP interface", as illustrated in Figure \ref{pcap}. This process is used to extract specific configuration frames from the entire bitstream, thereby enabling non-intrusive observation of the operational states of the DUT. 
Here, we use a Direct Memory Access (DMA) controller with AXI interconnection running at 100 MHz to guarantee an efficient data transfer from the configurable memory blocks in the DUT to the DDR memory onboard. 
During PCAP channel activation, the PL clock is suspended until the readback process concludes. This streamlined approach enables efficient control over the PL's operations as follows:

\begin{figure}[htbp]
\centerline{\includegraphics[scale=0.7]{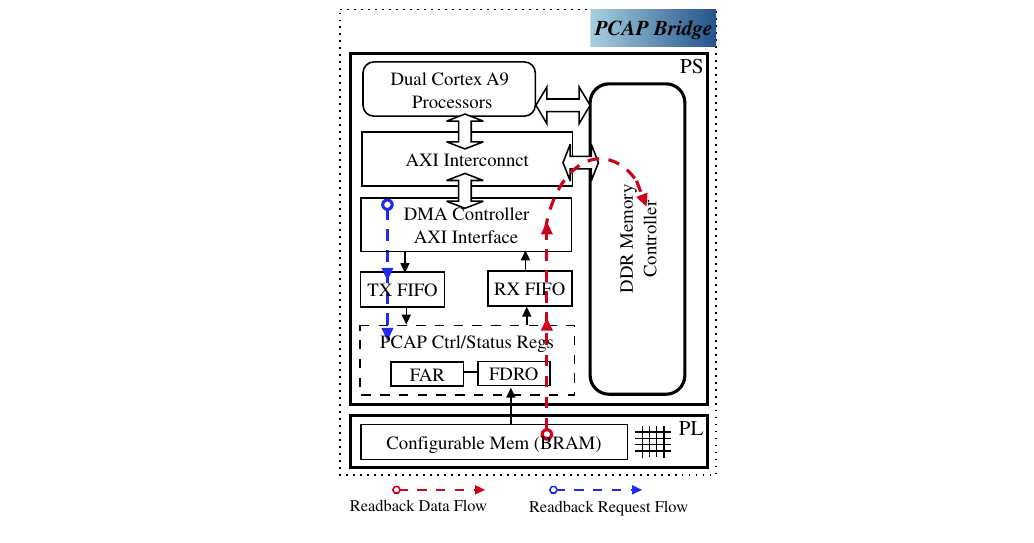}}
\caption{Data path and control path of PL readback via PCAP bridge.}
\Description{block diagrams of the data and control flow on FPGA}
\label{pcap}
\end{figure}

\begin{enumerate}
    \item The PCAP issues commands requesting data readback from the PL to the PS, then, DMA sends the commands to the PCAP Interface via AXI interface and the transmitter buffer (TX FIFO). The first frame address (FARD) to be read is set in the specific register FAR.
    \item Readback commands are acknowledged, and FARO fetches the target date frame, which is forwarded to the receiver buffer (RX-FIFO).
    \item The DMA controller transfers this data from the RX-FIFO to a designated location through the AXI interface.
\end{enumerate}


The key feature of this PCAP bridge is its ability to allow developers to modify target signals by altering frame addresses through the "PCAP Interface" unit rather than instantiating new signal probes in the RTL codes. 
The target frame addresses are specified in the "User Configuration" script. It acts as a static input to the "PCAP Interface" (as shown in Figure \ref{arbiter}) to keep track of the signals of interest. To achieve accurate tracing of the configurable memory cells in PL, we need to insert tracing information in the RTL design before the first build and implementation. The mechanism of locating the frame address is in three folds: 
\begin{itemize}
    \item \textbf{Block RAM Readback}. A read-only tracing mark (0xdeadbeef) is assigned to the first and the last addressable elements for the monitored BRAM blocks in the RTL code. Thus, the frame address of the memory block is captured when the tracing content is found in the readback frames. A similar method is applicable in our experiment for monitoring the General Purpose Registers (GPRs) implemented as BRAMs on FPGA.
    \item \textbf{Distributed RAM Readback}. To obtain the distributed LUTRAM contents, CAPTUREEE2 primitive \cite{amdpcap} is used in the RTL design for capturing the distributed memory cells. These values can then be read out of the device by reading configuration memory through the PCAP process. Register values are stored in the same memory cell that programs the register's initial state.
    \item \textbf{Placement constraints}. To enhance the tracing ability and make sure the contents show up in consecutive frames, we fixed the placement of the target memory cells in the XDC constraints. This applies to both BRAM and LUTRAM elements. Compared with behavioral modeling of the RAMs in RTL codes, structural modeling by instantiations of primitives requires less effort for tracing in the readback data.
\end{itemize}
\label{memreadback}

After porting the RTL design to the FPGA board, the extraction of the frame addresses is finished before the co-verification process starts. By adopting different frame addresses and frame lengths, the user is able to navigate to different monitored memory cells. This methodology significantly reduces re-implementation/re-compilation time and minimizes hardware resource utilization. Furthermore, the enhanced flexibility and scalability for PL observation provide advantages when exploring extensive design spaces in complex processor models. However, its effective utilization necessitates careful consideration of several key limitations:

\begin{itemize}
    \item The PCAP interface requires exclusive access to the configuration module. This requires that while the PCAP is active, other interfaces, such as JTAG, must be disabled. This restriction is necessary to ensure the integrity of the captured data and to prevent conflicts between different access methods.
    \item The PCAP readback operation is subject to data frame constraints. A single readback request cannot be fragmented across multiple DMA transfers (aligned to 4KB). This limitation arises from the fact that the readback process requires a fixed-size header and data frame of 101 words (404 Bytes) on Zynq7000 SoC, which includes the number of desired frames and an extra padding frame. As a result, the maximum data payload that can be transferred in a single transaction is limited to 10 frames (9 data frames + 1 padding frame), i.e., 4040 Bytes, meeting the requirement under 4096 Bytes. Attempting to read back more than 10 frames in a single request will inevitably lead to DMA transfer errors on Zynq 7000 SoCs.
\end{itemize}

To maximize the efficiency of PCAP readback, it is crucial to carefully plan the number of frames to be captured in each transaction. By adhering to the 10-frame limit, we can ensure reliable and error-free data transfer between the PL and PS.



\subsubsection{\textbf{Re-constructor} (block \textcolor{red}{\textcircled{3}} in Figure \ref{blockdiagram})} \label{rebuilder}
Proprietary solutions for RTL debugging and virtualization such as Siemens\textsuperscript{\textregistered} ModelSim or Xilinx XSim \cite{xsim} are widely adopted in the research and industry. In this work, we select a suite of lightweight open-source tools as alternatives to re-construct the cycle-accurate simulation due to their efficiency, flexibility, and significant advantages in processing time when verifying RTL designs. Cocotb \cite{cocotb} leverages Python’s asynchronous programming and libraries to create lightweight and scalable testbenches, reducing the complexity and overhead associated with traditional methodologies, e.g., Universal Verification Methodology (UVM) \cite{cocotb}. iCarus Verilog \cite{icarus}, as a lightweight simulator, is optimized for small to medium-sized designs, ensuring faster compile and simulation times compared to heavier commercial simulators that may introduce unnecessary latency for similar tasks \cite{icarus}. Finally, GTKWave \cite{gtkwave}, with its efficient handling of waveform data formats of Value Change Dump (VCD) file, provides a quick visualization of simulation results without the performance overhead of more complex, proprietary viewers \cite{gtkwave}. Notably, these tools minimize processing time while maintaining accuracy and reliability, making them ideal for iterative RTL design and verification in time-sensitive or resource-constrained environments.

The detailed functionalities are comprising:
\begin{itemize}
    \item \textit{Cocotb}, a Python-based framework, employs co-routines to facilitate lightweight concurrency. This design enables efficient synchronization between co-routines and the activities within an HDL simulation environment. In addition to the Python library, Cocotb utilizes native C/C++ libraries to integrate seamlessly with the simulation environment through simulator-specific APIs. 
    \item \textit{iCarus Verilog} serves as the Verilog simulator, compiling and executing the RTL designs. It provides a robust and efficient platform for digital circuit design and verification.
    \item \textit{GTKWave} completes the final step by providing waveform visualization capabilities. It offers a comprehensive set of features, including support for various waveform formats, hierarchical design exploration, signal filtering, and measurement capabilities. By employing GTKWave, researchers and engineers can effectively analyze the timing and logical behavior of their designs, facilitating the debugging and optimization process.
\end{itemize}

The reconstruction integrates seamlessly: Cocotb generates test vectors by parsing the ".json" checkpoint and interfaces with iCarus Verilog, which simulates the design and produces in the VCD files. Finally, GTKWave renders the dumped VCD and generates the waveforms. The key-value pairs within the ".json" checkpoint encapsulate metadata from the current check iteration. This metadata comprises the checkpoint ID, program counter, instruction mnemonic, and two sets of general-purpose register values—one extracted from the bitstream and the other from the instruction set simulator in the processing system.


\subsection{Typical Workflow} \label{workflow}
Figure \ref{work_flow} presents the simplified workflow orchestrating the aforementioned components, demonstrating their synergistic interaction in facilitating RTL design verification. This section introduces the typical workflow in the time dimension (stages), while the verification is also conducted in different spacial domains:
\begin{itemize}
    \item \textit{Host PC} handles the bitstream generation and workload insertion in the beginning and parses the checkpoint for visualization in waveforms.
        \item \textit{FPGA - ARM Processing System} executes ISS simulations, manages status checking and synchronization tasks, and leverages the efficient interconnect between PS and the PL for optimized data transfer.
    \item \textit{FPGA - Programmable Logic Area} performs the user-defined RTL logic (DUT) as a hardware emulator.
\end{itemize}

In the proposed workflow, the host PC is responsible for executing Stages 2 and 5. Especially, the PS on FPGA demonstrates extensive task involvement, actively incorporating with the host and PL in Stages 2, 3, and 4. The reconfigurable PL is dedicated to conducting the RTL emulation in Stages 2 and 3. The temporal stages are:

\begin{figure}[hbtp]
\centerline{\includegraphics[scale=0.63]{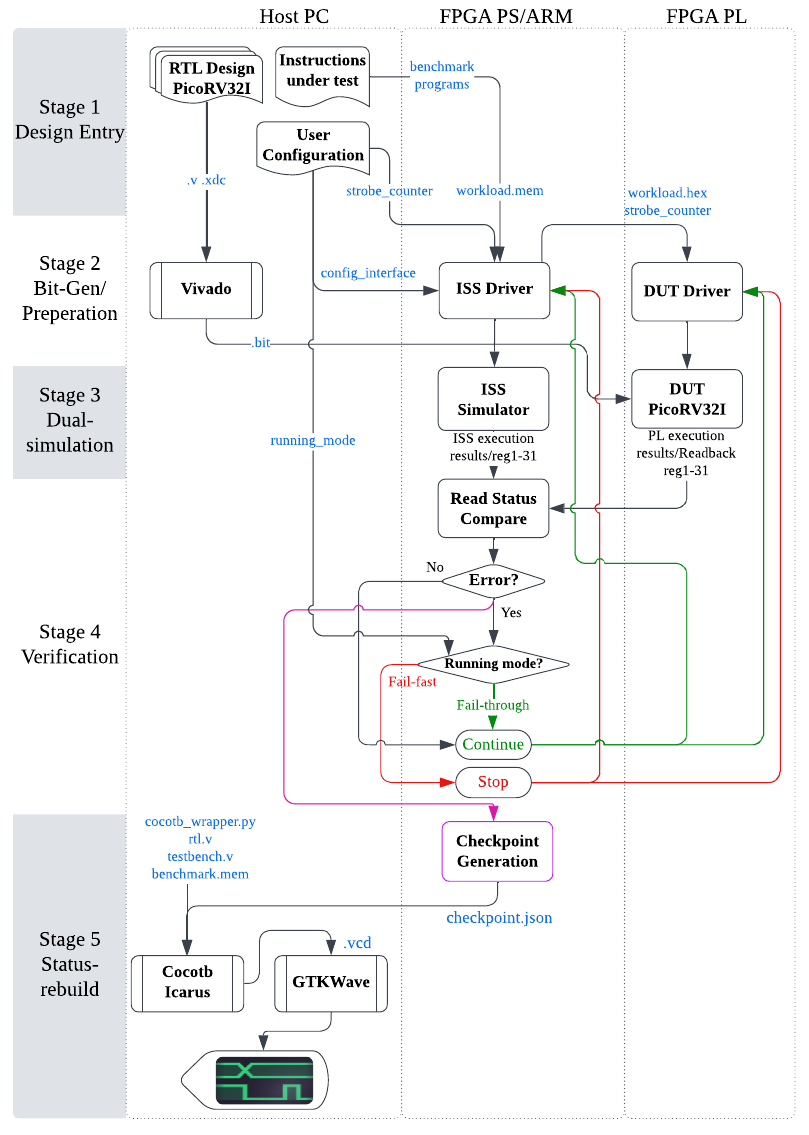}}
\caption{Typical workflow of FERIVer. The co-verification procedure is divided into 5 stages and processed in 3 domains.}
\Description{5 stages works flow of feriver}
\label{work_flow}
\end{figure}

\begin{figure*}[hbpt]
\centerline{\includegraphics[scale=0.8]{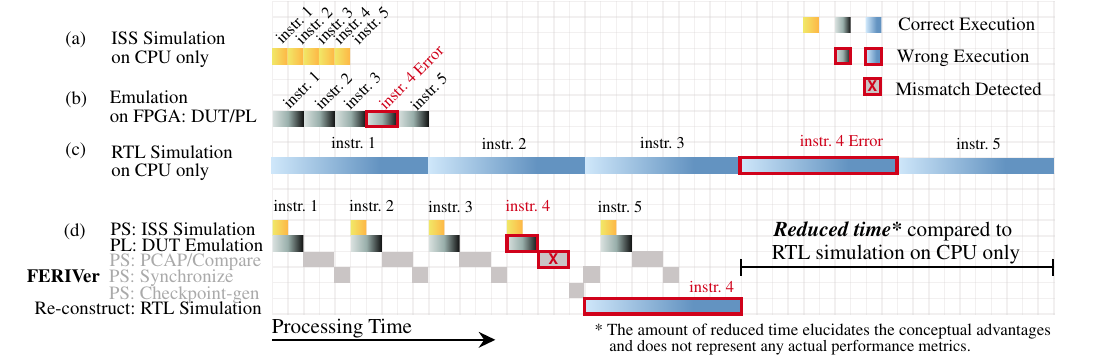}}
\caption{An illustrative comparison between conventional verification tools and the FERIVer method (absolute processing time, shorter is better). The example scenario presents the execution of 5 consecutive instructions, where the ISS employs an RV32I abstraction layer and works as a reference model, and the DUT entity fails on the execution of the 4th instruction.}
\Description{comparison of the four tools in terms of verification pipeline}
\label{timeallocates}
\end{figure*} 

\textbf{Stage 1} -- The necessary files of the RTL model are prepared to be processed with Xilinx Vivado tools. A configuration file is crucial at this stage, as it explicitly specifies the number of interleaved instruction checking (strobe\_counter) and configures the PCAP interface for reading the bitstream (config\_interface).

\textbf{Stage 2} -- The standard design flow of Vivado tools is applied to generate the hardware description file, which is then deployed on the FPGA. The benchmark programs are inserted into PS as workloads for evaluation purposes. The two drivers of ISS and DUT are responsible for single-step execution or skipping a certain number of instructions in every checking iteration, predefined in the strobe\_counter, i.e., pausing the execution and checking the results after every 3 instructions when strobe\_counter is set to 3.

\textbf{Stage 3} -- The software-based ISS in the PS and the hardware emulation in the PL execute the instructions concurrently. We prepare the same RV32I single-core workloads for the test. The ISS produces files containing the execution states, while the DUT in the PL provides the selected bitstream frames via the PCAP channel. Both ISS simulation and PL emulation pause with the toggling of the interrupt and resume only when the current checking results are available.

{\textbf{Stage 4} -- The "Arbiter" performs a comparison; any mismatch in the monitored registers triggers an error and halts the execution in both ISS and DUT, logging the discrepancy in the asynchronous buffers in the "Sync/Check" unit. The “Checkpoint Generator” structures the error, including register contents from both parties and PC of the current instruction, facilitating further RTL analysis in the next stage.

{\textbf{Stage 5} -- The waveform is generated using tools introduced in \href{rebuilder}{\S 3.1.3}, providing a human-friendly visualization for debugging.



\subsection{Mechanism of the Verification Acceleration} \label{sec-accl}

To illustrate our advantages in verification time compared to conventional verification processes, we visualize the time consumption of the different simulations in Figure \ref{timeallocates}. Firstly, note that the time used for the RTL code to generate the bitstream by Vivado is omitted because it only occurs once before the actual emulation starts. Secondly, the RTL designs of processors are sophisticated and typically exhibit errors in the execution during the verification process \cite{bugs}. Considering a scenario in a faulty processor design where 5 instructions are examined, the sub-figures of Figure \ref{timeallocates}, (a), (b), and (c) illustrate the sketchy time consumption by abstract simulation of ISS, hardware emulation on FPGA, and RTL simulation, respectively. The RTL simulation (c) with cycle-accurate analysis represents the most time-consuming process. 

Therefore, from a temporal perspective, the correct executions of the majority of instructions consume considerable simulation time. Figure \ref{timeallocates} (d) shows our approach, which targets to identify execution errors as early as possible by parallel simulation in both PL and PS. This allows for expedited verification of correct executions, thereby reducing overall verification time. As depicted in Figure \ref{timeallocates}, assuming an execution error rate of 25\% (1 out of 5 instructions), the FERIVer framework accelerates the overall verification process. The lower the error rate, the more complex the RTL design, and the more efficiency this approach can obtain. 
This hypothetical case only aims to demonstrate the potential of our approach. We present the comparison of verification speed for real benchmark scenarios in \hyperref[sec-evaluation]{\S 4.2}.

\section{Evaluation and Discussion}  \label{sec-evaluation}
We evaluate the performance of FERIVer and compare it with two CPU-based verification platforms: Xilinx XSim, and open-source Verilator \cite{verilator}. Xilinx XSim is an integral component of the Xilinx Vivado Design Suite, serving as an integrated Hardware Description Language (HDL) simulator for digital systems design and verification. Verilator distinguishes itself by compiling HDL design entities directly into high-level abstraction programming models, enabling rapid simulation speeds and seamless integration with testbenches.

\subsection{Overhead Analysis} \label{utilization}

The hardware usage of our experimental setup consists of two parts: the PicoRV32I processor and the infrastructure components for retrieving the data from PL. The Pico SoC project provides a size-optimized 32-bit RISC-V processor. Its simplified architecture eases our work of extracting the values of the execution state registers, which makes the Pico processor suitable as an object in our verification framework since it could be replaced by other RTL designs of any size. In the current prototype, we sampled 31 general-purpose registers in the Pico core. The FERIver framework occupies only 1.1\% (0.6K slices) of the programmable resources and 6.4\% block RAM on XC7Z020 FPGA, shown in Table \ref{hardware_utilization}.

\begin{table}[pbth]
\caption{Hardware utilization in PL.}
\begin{center}
\begin{tabular}{c|ccc}
\hline
\textbf{} & \textbf{LUT}& \textbf{BRAM36K}& \textbf{Flip-flop} \\
\hline
Available (Artix-7) & 53200 &140 &106400 \\
\hline
FERIVer & 475 (0.9\%)&9 (6.4\%) &124 (0.11\%)  \\
\hline
PicoRV32I (DUT) & 920 (1.7\%)&0 (0\%) &578 (0.54\%) \\
\hline
\end{tabular}
\label{hardware_utilization}
\end{center}
\end{table}

Running QEMU to simulate a RISC-V 32I ISA on the PS (ARM Cortex-A9 dual-core processor) introduces both hardware and software overhead. The QEMU and the RISC-V-related dependencies (libglib2.0, glibc, etc.) \cite{qemu} require approximately 650MB in the storage. At runtime, QEMU consumes around 48.8\% of the DDR3 (~ 250MB/512MB). The biggest performance impact comes from QEMU’s dynamic binary translation (DBT), introducing a translation overhead due to the single-threaded nature of QEMU's TCG (Tiny Code Generator), which heavily loads one ARM core while leaving the second core underutilized.

\subsection{Experimental Result and Analysis} \label{performance}




To compare with the existing CPU-based solutions for RISC-V ISA verification, we select three small-scale but representative workloads: skew sort (ssort), quick sort (qsort), and message digest algorithm 5 (md5). The testing programs are pre-compiled on the host and executed by the RISC-V processor models in both PS (ISS) and PL (RTL). The two baseline platforms (Verilator and Xilinx XSim) are both running on the host machine. Then we normalize the average execution speed to a million instructions per second, as shown in Figure \ref{123}. The observed verification speed of FERIVer is up to 5.31 MIPS measured when running the qsort benchmark. The overall performance presents a 20x-35x boost compared to the open-source Verilator and 150x-177x faster than the CPU-based Xilinx XSim platform.

\begin{figure}[hpbt]
\centerline{\includegraphics[scale=0.40]{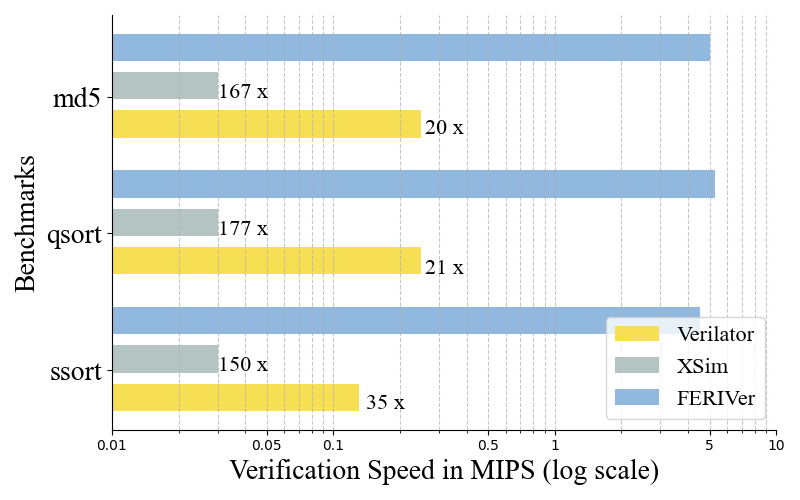}}
\caption{Comparison of verification speed (million instructions per second) on different platforms (longer is better).}
\Description{comp_speed}
\label{123}
\Description{Comparison of Verification Speed (Million Instructions per Second) on Different Platforms.}
\end{figure}


Since RTL designs often exhibit discrepancies compared to the expected execution results from the golden model, we intentionally introduce mismatches by modifying instructions in the "workload.mem" manually. This approach simulates a higher error rate of execution detected in this joint-verification method, reflecting the challenges encountered during RTL design and verification of RISC-V processors. As shown in Figure \ref{error_rate}, with the different error rates in the synthetic workload, our proposed method proves efficacious in accelerating the verification process. In the experiments, FERIVer still provided a 4x-32x speed-up to the other two verification tools even when encountering an error rate of 50\%.

\begin{figure}[hpbt]
\centerline{\includegraphics[scale=0.40]{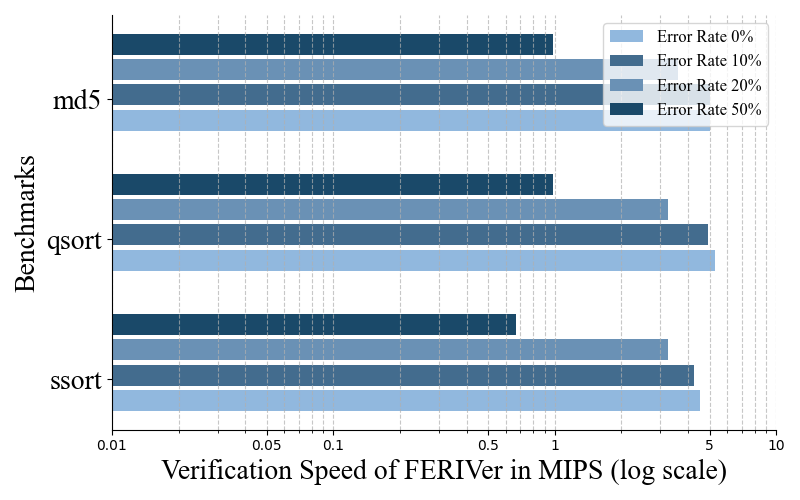}}
\caption{Verification speed of FERIVer on different error rates. The instruction execution errors are detected in RTL DUT, taking the ISS in PS as a golden model.)}
\label{error_rate}
\Description{Verification speed of FERIVer on different error rates in RTL DUT. The execution errors are introduced by synthetic erroneous workload.}
\end{figure}

The reduced speed primarily comes from the dominance of the time-consuming reconstruction process as the number of mismatches between ISS and RTL simulations increases. Besides, the original PCAP mechanism of Xilinx is able to encrypt the bitstream for security \cite{amdILA}. To obtain the least latency between PS and PL, our PCAP bridge runs in non-secure mode with a maximum bandwidth of approximately 140 MB/s, while the overall data transmission rate is limited by the PS AXI interconnect. In addition, the PL frames read back via PCAP contain a considerable amount of data for padding, therefore reducing the actual throughput.

\subsection{Related Work}



Recent works on non-intrusive verification have explored readback technology for FPGA-assisted emulation platforms. NIFD \cite{NIFD} facilitates single-step execution through a debugging interface with a bandwidth of 500 KB/s; however, it is insufficient for verifying many concurrent signals in complex RTL designs. GNOSIS \cite{gnosis} automates hardware execution verification and enables VCD file dumping for debugging but lacks support for FPGA on-chip memory readback. StateMover \cite{statemover_2020} achieves a debugging bandwidth of 100 MB/s but requires intricate configurations that adapt to specific data paths of different tasks. Besides, the substantial hardware utilization, 40K slices on Xilinx Ultrascale KCU105, is another limitation for further scaling. In contrast, FERIVer overcomes these limitations by offloading the primary validation tasks from the host CPU and managing critical communication between the RTL DUT and the reference model via a high-bandwidth channel (140 MB/s, \hyperref[performance]{\S 4.2}) between the PS and PL on the FPGA SoC, significantly enhancing verification efficiency while minimizing resource overhead (0.6K slices on Xilinx Zynq XC7Z020, \hyperref[sec-evaluation]{\S 4.1}).

\section{ Conclusion and Outlook} \label{sec-conclusion}

This work introduced FERIVer, an FPGA-assisted verification framework that is capable of cycle-accurate verification of RISC-V processors. It exhibits high efficiency by employing both hard IP processors and soft IP cores on AMD Xilinx SoC. Furthermore, we presented a dynamic and agile mechanism to parse checkpoints of the executions of the instructions, which can be applied to verify RISC-V or other computer architectures by simply replacing the abstract model and the RTL model under test. The evaluation results demonstrated the capability of significant time and effort savings for the verification of complex processor designs at very low costs. Our ongoing task focuses on optimizing the data throughput from PL to PS and verifying dedicated RISC-V ISA extensions for high-performance computation scenarios.

\bibliographystyle{ACM-Reference-Format}
\bibliography{ref}
\end{document}